\documentstyle[12pt]{article}
\begin{document}
\textwidth 15.5 cm
\textheight 25.9 cm
\topmargin -1.5cm
\pagestyle{empty}
\vspace{4.7  cm}

\noindent
{\bf THE WAVEFUNCTION RENORMALIZATION CONSTANT FOR THE
ONE- AND TWO-BAND HUBBARD HAMILTONIANS IN TWO DIMENSIONS}

\vspace{1.5  cm}

\hspace{2 cm} E. Louis,$^1$ G. Chiappe,$^1$ J. Gal\'an,$^2$
F. Guinea,$^2$

\hspace{2 cm} and J. A. Verg\'es$^2$

\vspace{0.5  cm}
\hspace{2 cm}$^1$Departamento de F{\'\i}sica Aplicada

\hspace{2 cm} Universidad de Alicante

\hspace{2 cm} Apartado 99, 03080 Alicante, Spain

\vspace{0.2 true  cm}
\hspace{2 cm}$^2$Instituto de Ciencia de Materiales

\hspace{2 cm} Consejo Superior de Investigaciones Cient{\'\i}ficas

\hspace{2 cm} Cantoblanco, 28049-Madrid, Spain

\vspace{1  cm}
\noindent{\bf INTRODUCTION}
\vspace{0.5 cm}

The normal state of the recently discovered high-$T_c$ superconductors
\cite{1}, exhibits many exotic properties that are not yet fully
understood. A picture that is gaining acceptance assumes that the normal
state is not a normal Fermi liquid [2-5] but rather it shows
unconventional properties somewhat similar to those of the Hubbard model
in 1D \cite{6}, namely: i) an unrenormalizable Fermi-surface phase
shift, which implies a vanishing wavefunction renormalization constant
(Z), ii) a one-particle spectral density that shows a strong peak at the
Fermi energy, iii) a Fermi surface that obeys Luttinger's theorem
\cite{7}, and, iv) charge and spin separation.

Only in very few cases has a quantitative check of these properties
been undertaken [8-15]. In particular, Z has
been calculated by exactly solving either the t-J \cite{12} or
the Hubbard Hamiltonian in $4 \times 4$ clusters. Charge and spin
separation has also been recently investigated on finite clusters of the
square lattice \cite{13}. Unfortunately the behavior of these properties
in the thermodynamic limit cannot be investigated at present by means of
exact calculations.

In previous papers \cite{16} we have presented an Unrestricted
Hartree-Fock (UHF) calculation of Z for the one-band Hubbard model in
the square lattice. UHF allows to consider clusters large enough to
investigate the scaling of Z with the system size. The results indicate
that the system shows non-conventional behavior (Z=0) near half-filling.
For a given value of U, Z vanishes at low dopings, and becomes finite as
doping is increased. For very large U the value of Z remains finite for
all dopings, excluding half-filling. In this work we discuss these
results outlining the reasons why we think that UHF may provide
trustable information on Z. We also show that the results are not
changed if the wavefunction is written as a linear combination of UHF
solutions each one centered on a cluster site (this procedure
significantly improves the UHF results for the total energy). Finally we
present a calculation of Z for the two-band Hubbard model; again, near
half-filling, Z scales to zero with the system size.

\vspace{0.5 cm}
\noindent{\bf THE HUBBARD HAMILTONIAN AND THE UHF APPROXIMATION}
\vspace{0.5 cm}

The Hubbard Hamiltonian for one and two bands can be written in the
general form \cite{17},

\begin{equation}
H = H_{0} + H_{1},  \\
\end{equation}

\begin{equation}
H_{0} = \sum_{i\sigma} E_{i} n_{i\sigma}
-\sum_{<ij>\sigma} t_{i,j} c_{i\sigma}^{\dagger}c_{j\sigma}, \\
\end{equation}

\begin{equation}
H_{1} =  \sum_{i} U_{i}n_{i\uparrow}n_{i\downarrow}
\end{equation}

\noindent
where the indexes i,j run over the atoms of the CuO$_2$ planes in the
two-band case, or over atoms at sites of the square lattice in the
one-band case. The operator $c_{i\sigma}$ destroys an electron with a
z-component of the spin $\sigma$ at orbital i, and $t_{i,j}$ is the
hopping matrix element between orbitals located at atoms $i$ and $j$
(the symbol $<ij>$ denotes the sum over all pairs of nearest neighbors),
$E_{i}$ are the orbital energies and $U_{i}$ is the intrasite Coulomb
repulsion.

The most general effective Hamiltonian, within the UHF approximation,
can be written as \cite{18,19},

\begin{eqnarray}
H_{1}^{eff}=\sum_{i}(c^{\dagger}_{i\uparrow},c^{\dagger}_{i\downarrow})
\left( \begin{array}{cc} <n_{i\downarrow}> &
-<c^{\dagger}_{i\uparrow}c_{i\downarrow}>^{*} \\
-<c^{\dagger}_{i\uparrow}c_{i\downarrow}> & <n_{i\uparrow}> \end{array}
\right) \left( \begin{array}{c} c_{i\uparrow} \\ c_{i\downarrow}
\end{array} \right) \nonumber \\
-\sum_{i} U_{i} (<n_{i\uparrow}><n_{i\downarrow}> - \left| <c_{i\uparrow}
^{\dagger}c_{i\downarrow}> \right| ^{2})
\end{eqnarray}

\noindent
If spin flip terms characterized by a non-zero value of
$<c^{\dagger}_{i\uparrow}c_{i\downarrow}>$ are ignored, the standard
Hartree-Fock approximation of the Hubbard interaction is recovered. In
this work we have only considered solutions with a single non-zero
component of the local magnetization for the following reasons: i) near
half-filling the vortex solutions described in \cite{18,19} lie at much
higher energies than the Ising solutions (magnetic polaron or extended),
and, ii) at moderate dopings vortices become more competitive but only
for an even number of particles. In general, solutions with transverse
magnetization are more extended, and lead to lower values of Z.

\newpage
\vspace{0.5 cm}
\noindent{\bf THE WAVEFUNCTION RENORMALIZATION CONSTANT}
\vspace{0.5 cm}

 The wave-function renormalization constant is given by

 \begin{equation}
\sqrt{Z} ={{ <\Psi(N)|c_{\alpha}\Psi_{\alpha}(N+1)> } \over
{<\Psi(N+1)|c^{\dagger}_{\alpha}c_{\alpha}| \Psi(N+1)>}}
\end{equation}

\noindent
where $\alpha$ stands for the appropiate quantum numbers and the
$\Psi$'s are the ground state wave-functions. In calculating Z we
proceed as follows. We start from the Slater determinant for N electrons
and empty the one-electron state of highest energy. This is equivalent
to identify the operator $c_{\alpha}^{\dagger}$ in Eq. (2) with

\begin{equation}
c_{E_n\sigma}^{\dagger} = \sum_{\bf{k}}\gamma_{\bf{k}\sigma}
(E_n)c_{\bf{k}\sigma}^{\dagger},
\end{equation}

\noindent
where $E_n$ and $\sigma$ are the energy and spin of the empty UHF level
of lowest energy, and $\gamma_{\bf{k}\sigma}(E_n)$ are coefficients
obtained in the UHF calculation. For this choice of the operators in Eq.
(5) the normalization constant in this equation is equal to one.
Simultaneously, we introduce a small random distortion (maximum modulus
of $10^{-6}$) on the starting mean field, both on local charges and
magnetizations. The addition of this distortion has two objectives: a)
to break the degeneracy of the filled state of highest energy as occurs
in cases such as half-filling, and, b) to draw the iteration process
towards the state of lowest energy. Then we initiate the iteration
process \cite{18,19} until selfconsistency is achieved.

The square root of the wave-function renormalization constant Z is
given by the overlap between the resulting Slater determinant and the
initial one, which can be calculated through the following expression,

\begin{equation} <\Psi(N)|\Phi(N)> = \left| \begin{array}{ccc}
<\psi_{1}|\phi_{1}> & \cdots & <\psi_{1}|\phi_{N}> \\ . & \ldots & . \\
<\psi_{N}|\phi_{1}> & \cdots & <\psi_{N}|\phi_{N}> \end{array} \right|
\end{equation}

\noindent
where $\Psi(N)$ and $\Phi(N)$ are two Slater determinants for N
particles, and $\psi_{i}$ and $\phi_{i}$ the corresponding
monoelectronic wavefunctions.

\vspace{0.5 cm}
\noindent{\bf One-band Hamiltonian}
\vspace{0.5 cm}

The one-band Hubbard Hamiltonian has been solved on finite clusters of
size $L \times L$ (with L up to 16). For the parameters in Eqs. (1-3) we
took, $t_{i,j} = 1$, $E_{i} = 0$ and $U_{i} = U$. The ground state
for intermediate values of U is, in the case of a single hole, the
magnetic polaron, whereas for many holes it corresponds to magnetic
polarons disorderly distributed through the whole cluster. At $U=\infty$
the UHF ground state is ferromagnetic for all fillings \cite{16}.

\newpage
\vspace{10cm}
Figure 1: UHF results for the wave-function renormalization constant
Z vs the cluster size ($L \times L$) for the one-band Hamiltonian and
$6.25\%$ doping.
\vspace{0.3 cm}

Our results for Z at half-filling clearly indicate that it slowly
decreases with L for all U. The UHF results for the $4 \times 4$ cluster
are in good agreement with the exact results reported in Refs.
\cite{10,12}. The scaling of Z to zero with the system size, can be
understood as follows. The changes in the charge and spin configurations
that result from emptying one of the degenerate levels at the top of the
lower Hubbard band are non-uniformly localised along lines. Thus, as the
one-hole selconsistent solutions are localized, it can be expected that
the overlap between the two should decrease with the size of the
cluster. It should be noted that as at half-filling the system is an
antiferromagnetic insulator, strictly speaking it does not make sense
to talk about its Fermi or non-Fermi like character. We note, however,
that homogeneous HF solutions would have led to a finite Z even at
half-filling. Thus, the present results clearly indicates that for a
small, still finite, density of holes the system should show
non-conventional behavior.

\vspace{10cm}
Figure 2: Schematic phase diagram for the one-band Hubbard model
according to the results for Z. The parameters are the fraction of holes
referred to half-filling (x) and U. Z vanishes in a region close to
half-filling.
\vspace{0.3 cm}

As regards finite doping, we have analyzed the case of $6.25\%$ holes.
The results of Figure 1 indicate that, for $6.25\%$ of holes, Z scales
to zero for U = 20 whereas it remains finite for U = 10. This defines a
transition from Fermi- to Luttinger-like behavior at an intermediate U.
Thus, we may expect that at large enough U the constant Z will again be
finite. At $U=\infty$ the UHF results show that Z remains finite at all
dopings, except at half filling (this is a consequence of the
ferromagnetic character of the UHF ground state away from half-filling).
On the other hand, in the dilute limit the results of Ref. \cite{20}
indicate that Z is finite for all U.

The numerical data for Z, in the cases for which it decreases with L, can
be fitted by means of the expression \cite{16}

\begin{equation}
\sqrt{Z}=a \exp (-b \ln L)
\end{equation}

This equation coincides with that suggested years ago by Anderson
\cite{2,21}, and indicates that there exists an unrenormalizable
phase shift at the Fermi energy. Actual results for b show appreciable
errors due to the difficulty of the calculations. In all cases
analysed here, $b\approx0.5$. The resulting phase shift $\delta$,
obtained by comparing Eq. (8) with that in refs. \cite{2,21}, is
$\delta\approx \pi/\sqrt{2}$, or by introducing the Fermi wave number,
$\delta\approx0.9k_{F}a$, in excellent agreement with the result of
\cite{2}.

The previous results suggest the phase diagram for the one-band Hubbard
model depicted in Figure 2. The region where the system is expected to
show non-conventional behavior is close to half filling and finite U.
For a given finite U, the system will first (low doping) be a Luttinger
liquid and become a Fermi liquid as doping is increased.

\vspace{0.5 cm}
\noindent{\bf Two-band Hamiltonian}
\vspace{0.5 cm}

We have chosen the following set of parameters, $E_p-E_d = 4.0$,
$U_d=6.0$, $t_{pd}=1.0$, and the remaining parameters equal to zero.
This choice gives, for instance, a reasonable fit of the experimental
data for the magnetization. Clusters containing up to $12 \times 12$
unit cells, have been considered. Both electron and hole dopings have
been investigated. For a single hole the ground state is, in both cases,
the magnetic polaron \cite{19}. The results for Z at half-filling are
shown in Figure 3. As in the case of the one-band Hamiltonian, Z
decreases with L as in Eq. (8). The scaling of Z to zero is very
similar for both types of doping.

\vspace{10cm}
Figure 3: UHF results for the wave-function renormalization constant
Z vs the cluster size ($L \times L$) for the two-band Hamiltonian at
half-filling. Results for both hole and electron doping are shown.

\newpage
\vspace{0.5 cm}
\noindent{\bf BEYOND UHF}
\vspace{0.5 cm}

The magnetic polarons, although show the appealing features of all mean
field solutions \cite{18,19}, have many drawbacks. The two most
outstanding are the non-uniform distribution of the excess charge
[13-15], in contrast with the exact solutions [5-9],
and an energy, significantly higher than the exact one \cite{15}. In
particular the former could be the origin of the behavior of Z discussed
above. The usual way followed in quantum chemical calculations to go
beyond HF is the so-called Configurations Interaction (CI). In the case
of the magnetic polarons a particular CI seems rather clear, namely, the
one that allows the localised polarons to move through the whole
cluster. We have explored this CI for a single hole, by writting its
wavefunction as a linear combination of Slater determinants for the
single polarons (SP) localised on all cluster sites. The exact
Hamiltonian has been solved in this basis set. We call this solution a
multipolaron (MP).

Thus, we write the wavefunction of a single hole ($N_{s}-1$ electrons) as
a linear combination of single polaron Slater determinants centered on
all cluster sites

\begin{equation}
\Psi_{MP}(N_{s}-1)=\sum_{i}a_{i}\Phi^{i}(N_{s}-1)
\end{equation}

where MP stands for multipolaron, and $\Phi^{i}(N_{s}-1)$ is the
wavefunction for the magnetic polaron centered on site i. The
coefficients $a_{i}$ are obtained through diagonalization of the exact
Hamiltonian.

The ground state wavefunction for a single hole includes all the SP
wavefunctions with the same weight, and, therefore, gives an excess
charge uniformly distributed throughout the whole cluster, as in the
exact case. Note, however, that the $S_z$ component of the single
polaron ($S_z = 1/2$) is preserved by the multipolaron approximation.
The results for the energy necessary to create one hole (the difference
between the energies of the system with one and zero holes) in the SP
and MP approximations, are reported in Table I. As expected, the MP
approximation lowers the energy of the single polaron. The most
important changes are found for intermediate values of U. The correction
rapidly decreases with U, and tends to zero for infinite U. On the other
hand we note that the correction does only slightly depend on the size
of the cluster. This can be easily understood by noting that the
interaction between polarons rapidly decreases with the distance.

We note that although the MP solution restores the translational
symmetry of the Hubbard hamiltonian, features associated with the
initial, localized solutions, remain; in particular the behavior of Z is
not changed. This is so because, in our CI scheme, each of the solutions
which enter in the final wavefunction contains a "continuum" of levels
which is substantially modified by the addition of an extra electron.
The orthogonality catastrophe which is responsible for the absence of
coherent quasiparticles within the Hartree-Fock approximation is
unchanged in the final wavefunction.

\vspace{0.5 cm}

\begin{table}
\caption{Energy (in units of t) required to create a hole in the single
polaron (SP) and multipolaron (MP) approximations, for three cluster sizes
(L) and several values of the intrasite Coulomb repulsion (U). Exact
results are taken from Refs. [7,8]. The HF solution for the
$4 \times 4$ cluster and $U=4$ is not the single polaron (see text).}
\vspace{0.3 cm}

\begin{tabular}{cclccclccc}
\hline
\hline
   &       & & &    & L$\times$L & &  &  &  \\
   & exact & & & SP & & &  & MP &   \\
\cline{2-2} \cline{4-6} \cline{8-10}
U  & $4\times4$ & & $4\times4$ & $8\times8$ & $12\times12$ & &
$4\times4$ & $8\times8$ & $12\times12$ \\
\hline
    4 & -1.0434 &&  -0.7352 & -0.7317 & -0.7270 & & -0.7942 & -0.9219 & -0.9366
\\
    6 & -1.4148 &&  -0.8956 & -0.9031 & -0.9033 & & -1.2143 & -1.2413 & -1.2419
\\
    8 & -1.6784 &&  -1.0719 & -1.0793 & -1.0793 & & -1.3731 & -1.3940 & -1.3941
\\
   10 & -1.8640 &&  -1.2092 & -1.2190 & -1.2190 & & -1.4662 & -1.4868 & -1.4868
\\
   12 & -2.0015 &&  -1.3146 & -1.3267 & -1.3266 & & -1.5326 & -1.5527 & -1.5527
\\
   16 & -2.1954 &&  -1.4627 & -1.4767 & -1.4767 & & -1.6249 & -1.6432 & -1.6432
\\
   20 & -2.3281 &&  -1.5601 & -1.5741 & -1.5741 & & -1.6871 & -1.7034 & -1.7034
\\
\hline
\hline
\end{tabular}
\end{table}

\vspace{0.5cm}

\vspace{0.5 cm}
\noindent{\bf CONCLUDING REMARKS}
\vspace{0.5 cm}

We have presented the first investigation of the scaling of the
wave-function renormalization constant for the one-band Hubbard
Hamiltonian in the square lattice. Our results suggest that it may show
non-conventional behavior in the thermodynamic limit, as discussed by
previous authors [2-5]. The region of the parameter space (U
and filling) where Z vanishes is narrow and close to half-filling. The
way in which Z scales with the size of the system makes this problem
very difficult to investigate by means of exact calculations.

It is worth mentioning that our route towards the results which show
highly unconventional behavior is itself very conventional. We apply the
most standard technique in the study of many-body systems: the
Hartree-Fock approximation. While the method, and the results, seem to
be a contradiction of terms, we think that the opposite is true.
Precisely because the method is so standard, it is biased in favor of
normal (Fermi liquid) behavior. In fact, we recover such behavior in the
low density limit, as we should. Also, the AF insulator at half filling
is well described within our approximation. The main deviation of our
results from conventional behavior is the proliferation of solutions
which break translational symmetry, near half-filling. It is also in
that region where we find Z=0. Restoring the translational symmetry
by means of a CI scheme, while improving the results for the total
energy, does not change the results for Z.

Some results for the two-band Hamiltonian, indicate that this model may
also show non-conventional behavior, no matter the type of doping.

\vspace{0.5 cm}
\noindent{\bf ACKNOWLEDGEMENTS}
\vspace{0.5 true cm}

The financial support from the spanish CICYT (grant MAT91-0905-C02) is
gratefully acknowledged. One of us (G. Chiappe) wishes to thank the
"Ministerio de Educaci\'on y Ciencia" for a postdoctoral fellowship
and the "Fundaci\'on Gil-Albert for a partial grant.


\newpage
\begin{thebibliography}{99}

\bibitem{1}
J. G. Bednorz and K.A. Muller,
Z. Phys.\ B {\bf 64}, 88 (1986).

\bibitem{2}
P.W. Anderson,
Science \ {\bf 235}, 1196 (1987);
ibid, Phys.\ Rev.\ Lett.\ {\bf 65}, 2306 (1990);
ibid, {\bf 64}, 1839 (1990).

\bibitem{3}
N. Nagaosa and P.A. Lee,
Phys.\ Rev.\ Lett.\ {\bf 64}, 2450 (1990).

\bibitem{4}
L.B. Yoffe and P.B. Wiegmann,
Phys.\ Rev.\ Lett.\ {\bf 65}, 653 (1990).

\bibitem{5}
C.M. Varma, P.B. Littlewood, S. Schmitt-Rink, E. Abrahams and A.E.
Ruckenstein,
Phys.\ Rev.\ Lett.\ {\bf 63}, 1996 (1989).

\bibitem{6}
F.D.M. Haldane,
J.\ Phys.\ C {\bf 14}, 2585 (1981).

\bibitem{7}
J.M. Luttinger,
Phys.\ Rev.\ {\bf 121}, 942 (1961).

\bibitem{8}
X.G. Wen,
Phys.\ Rev.\ B {\bf42}, 6623 (1990).

\bibitem{9}

\bibitem{10}
G. Fano, F. Ortolani and A. Parola,
Phys.\ Rev.\ B {\bf46}, 1048 (1992).

\bibitem{11}
H.E. Castillo and C.A. Balseiro,
Phys.\ Rev.\ Lett.\ {\bf 68}, 121 (1992).

\bibitem{12}
E. Dagotto and J.R. Schrieffer,
Phys.\ Rev.\ B {\bf43}, 8705 (1991).

\bibitem{13}
E.A. Jagla, K. Hallberg and C.A. Balseiro,
Phys.\ Rev.\ B{\bf 47}, 5849 (1993).

\bibitem{14}
S. Sorella, this volume.

\bibitem{15}
W. Metzner, this volume.

\bibitem{16}
G. Gal\'an, G.Chiappe, E.Louis, F.Guinea y J.A.Verg\'es,
Phys.Rev.B {\bf 46}, 3163 (1992);
E. Louis, G. Chiappe, J. Gal\'an, F. Guinea and J.A. Verg\'es,
Phys.\ Rev.\ B, {\bf 48}, 426 (1993).

\bibitem{17}
J. Hubbard, Proc. Roy. Soc. (London) Ser. A {\bf 276}, 238 (1963)

\bibitem{18}
A.R.Bishop, F.Guinea, P.S.Lomdahl, E.Louis y J.A.Verg\'es,
Europhys. Lett. {\bf 14}, 157 (1991);
J.A.Verg\'es, E.Louis, P.S.Lomdahl, F.Guinea y A.R.Bishop,
Phys.Rev.B {\bf 43}, 6099 (1991).

\bibitem{19}
J.A. Verg\'es, F. Guinea and E. Louis,
Phys.\ Rev.\ B {\bf 46}, 3562 (1992).

\bibitem{20}
J.R. Engelbrecht and M. Randeria,
Phys.\ Rev.\ Lett.\ {\bf 65}, 1032 (1990);
ibid, Phys.\ Rev.\ Lett.\ {\bf 66}, 3225 (1991.

\bibitem{21}
P.W. Anderson,
Phys.\ Rev.\ {\bf 164}, 352 (1967).

\end{thebibliography}
\end{document}